# Electric field effects on magnetotransport properties of multiferroic Py/$YMnO_3$/Pt heterostructures

V. Laukhin[1,2], X. Martí[1], V. Skumryev[2,3], D. Hrabovsky[1], F. Sánchez[1],
M.V. García-Cuenca[4], C. Ferrater[4], M. Varela[4], U. Lüders[5], J.F. Bobo[5],
and J. Fontcuberta[1]

[1] Institut de Ciència de Materials de Barcelona-CSIC, Campus UAB, Bellaterra 08193, Spain

[2] Institut Català de Recerca i Estudis Avançats (ICREA), Barcelona, Spain

[3] Departament de Física, Universitat Autònoma de Barcelona, Bellaterra 08193, Spain

[4] Departament de Física Aplicada i Òptica, Universitat de Barcelona, Diagonal 647, Barcelona 08028, Spain

[5] LNMH ONERA-CNRS, BP 4025, 31055 Toulouse, Cedex 4, France

Abstract

We report on the exchange bias between antiferromagnetic and ferroelectric hexagonal $YMnO_3$ epitaxial thin films sandwiched between a metallic electrode (Pt) and a soft ferromagnetic layer (Py). Anisotropic magnetoresistance measurements are performed to monitor the presence of an exchange bias field. When the heteroestructure is biased by an electric field, it turns out that the exchange bias field is suppressed. We discuss the dependence of the observed effect on the amplitude and polarity of the electric field. Particular attention is devoted to the role of current leakage across the ferroelectric layer.

## Introduction

Magnetic exchange bias between antiferromagnets (AF) and soft ferromagnets is being used to pin the magnetization of one of the magnetic layers in giant magnetoresistance and magnetic tunnel junction devices. In these devices, the magnetic state of the free magnetic layer is typically controlled by the oersted magnetic field created by suitable current lines. However, the unavoidable Joule heating at the current lines constitutes a severe limitation towards further integration of magnetic tunnel junctions and thus research must be focused to overcome this bottleneck. One of the possible alternatives is to switch from current to voltage control of magnetic devices. Biferroic materials, such as those displaying coupled ferromagnetic (FM) and ferroelectric (FE) behavior, could be employed for this purpose. However, magnetoelectric coupling in these materials is weak and magnetic switching of polarization has been demonstrated at low temperatures [1]. Alternatively, switching of magnetization could also be achieved, in principle, by using a device in which the exchange bias is controlled by an electric field [2]. This requires AF materials, in which the magnetic state can be controlled by an electric field. Fortunately, there are several AF materials that display simultaneously a ferroelectric (FE) character and in which the antiferromagnetic domains are coupled to the ferroelectric ones [3]. In principle, under these circumstances, application of an electric field would allow to simultaneously modify the ferroelectric and antiferromagnetic domains. As magnetic exchange bias between antiferromagnets and ferromagnets is intimately related to the domain structure of the antiferromagnet, it turns out that one could expect to tune exchange bias by an electric field.

This approach was early explored using magnetoelectric, but not multiferroic, $Cr_2O_3$ single crystals and a soft ferromagnetic upper grown layer [4]. Although $Cr_2O_3$ is not ferroelectric, it has been shown [4] that non-volatile exchange bias response can be achieved by proper magnetoelectric field cooling in simultaneously applied parallel and antiparallel magnetic and electric field.

The hexagonal phase of $YMnO_3$ is ferroelectric below ~ 800 K and it displays antiferromagnetic order below ~ 80K. Therefore at low temperatures $YMnO_3$ is a biferroic oxide. Moreover, it has been proved that the magnetic and ferroelectric domains are coupled [3]. Exploiting these characteristics in epitaxial thin films we have been able to induce an exchange bias between a $YMnO_3$ thin layer -grown on a bottom metallic Pt electrode- and an upper grown soft ferromagnetic Permalloy layer (Py) [5].

Exchange bias has been evidenced by the existence of a clear shift of the magnetization loop when the magnetization of Py is measured after appropriate field-cooling (FC) conditions [5]. Moreover, when a biasing voltage is applied between the Pt and Py electrodes, across the FE $YMnO_3$ film, it has been found that the exchange bias is rapidly suppressed. Remarkably enough, under appropriate experimental conditions, it has been shown that the magnetization (**M**) of Py is switched by application of an electric field [6].

The presence of an exchange bias has also been monitored by measuring the anisotropic magnetoresistance (AMR) of the Py film. The presence of an exchange bias field induces a distinct angular dependence of the AMR [7]. Exchange bias at the interface between ferromagnetic and antiferromagnetic materials is recognized to be associated to the development of a unidirectional magnetic anisotropy that pins the magnetization of an upper-grown ferromagnetic layer. As a consequence, when a magnetic field is applied parallel to the interface, the magnetization of the ferromagnetic layer does not follow (neglecting the anisotropy of the FM layer) the external field $\mathbf{H_a}$ but the $\mathbf{H_a}$+$\mathbf{H_{eb}}$ vector sum, where $\mathbf{H_{eb}}$ is the exchange bias field. The presence of $\mathbf{H_{eb}}$ affects the angular dependence of AMR of the FM layer when the external magnetic field is rotated [5-7]. If $\theta_M$ is the angle between the magnetization (**M**) and the measuring electric current direction (**J**), the resistivity is given by $\rho(\theta_M) = \rho_\perp + \Delta\rho\cos^2\theta_M$, where $\Delta\rho \equiv \rho_{//} - \rho_\perp$ and $\rho_\perp$ and $\rho_{//}$ are the resistivity for **M**⊥**J** and **M**//**J,** respectively. In presence of $H_{eb}$, when rotating $\mathbf{H_a}$ to an angle $\theta_a$ with respect to **J**, the measured $\rho(\theta_a)$, in general, does not follow a simple quadratic $\cos^2(\theta_a)$ dependence. Thus departure from this simple behavior allows monitoring the existence of $H_{eb}$ and its eventual modifications under a biasing electric field. The field-cooling conditions of the Py/$YMnO_3$/Pt heterostructure, and more precisely the angle ($\theta_{FC}$) formed by the measuring current **J** direction and the cooling field direction, are determining the particular shape of the angular dependence of the AMR.

Here, we report and compare the AMR data collected using different cooling conditions and the effects of electric field on magnetic exchange bias. Particular emphasis will be devoted to the role of current leakage across the $YMnO_3$ layer on the measured response.

## Experimental

$YMnO_3$ (0001) films, 90 nm thick, with hexagonal structure were grown by pulsed laser deposition on $SrTiO_3$ (111) substrates buffered with a thin epitaxial Pt layer (8 nm) as bottom metallic electrode. This heterostructure was covered by a Py film (15 nm). The metallic bottom Pt and top Py electrodes were grown by sputtering. X-ray diffraction experiments indicated that the Pt and $YMnO_3$ films are epitaxial, with (111) and c-axis out-of-plane orientation, respectively. Extensive structural details will be reported elsewhere [8]. During the growth of $YMnO_3$, a mask was used -partially covering the bottom Pt electrode- for subsequent electric contacting. Four (in-line) electric contacts (~$1.75\times0.2$ mm$^2$) on Py were used for transport measurements. We denote by $R_{Py} \equiv V_{Py}/I_{Py}$ the resistance measured between two $V_{Py}$ contacts, where $V_{Py}$ is the measured voltage and $I_{Py}$ (100 μA) is the injected current. Additional electrical contacts on Py and Pt were made for electric biasing ($V_e$) of the Py/$YMnO_3$/Pt sandwich (Fig. 1-top). The room-temperature resistivity of the $YMnO_3$ layer is of about $10^6$ Ωcm. AMR was measured by rotating an external field $\mathbf{H_a}$ in the plane of the sample (Fig. 1-bottom) by using a PPMS from Quantum Design.

## Results and discussion

AMR measurements of Py were performed after field-cooling (FC) the sample from room-temperature with the magnetic field (3 kOe) applied at the angle $\theta_{FC} = 0^o$ or $\theta_{FC} = 45^o$ from the measuring current direction. Once at the targeted temperature, the magnetic field is reduced down to the measuring field $H_a$ value, and it is afterwards rotated clockwise from -5º to 365º and anti-clockwise to the original position ($\theta_a = 0^o$ corresponds to $\mathbf{J}//\mathbf{H_a}$) while monitoring the voltage drop $V_{Py}$ along the Py track thus measuring its resistance as a function of the rotating angle, $R_{Py}(\theta_a)$.

Fig. 2 collects the clockwise $R_{Py}$ ($\theta_a$) curves at 5 K and $H_a$ = 50 Oe, obtained after FC at $\theta_{FC} = 0^o$ (curve a) as well as at $\theta_{FC} =45^o$ at 5 K ($H_a$ = 50 Oe) (b), 50 K ($H_a$ = 40 Oe) (c), and 100 K ($H_a$ = 50 Oe) (d). We first notice that at 100 K (Fig. 2-d), $R(\theta_a)$ is well described by a $\cos^2\theta_a$ dependence which implies that at this temperature $\theta_a = \theta_M$ and thus $\mathbf{M}$ is parallel to $\mathbf{H_a}$. Indeed, this is what should be expected as $H_{eb}$ is zero in the paramagnetic phase of $YMnO_3$ and thus any trace of exchange bias is absent. Therefore, observation of the $\cos^2\theta$ dependence is a clear signature of absence of $H_{eb}$. As expected, when reducing the temperature, due to the increase of $H_{eb}$, the $\cos^2\theta_a$ dependence gradually disappears. It is important to notice at 50 K, as shown in Fig. 2-c, effect of $H_{eb}$ is visible. At 5K the $\cos^2\theta$ dependence has disappeared absolutely, only one minimum in $R_{Py}(\theta_a)$, occurring at about $\theta_a \sim 90^o$ for $\theta_{FC} = 0^o$ (Fig. 2-a) or $\theta_a \sim 135^o$ for $\theta_{FC} = 45^o$ (Fig. 2-b) remains visible. This indicates that at this temperature, under a rotating field ($H_a$ = 40 Oe), the film magnetization remains pinned along the $\mathbf{H_{eb}}$ direction [7]. The observed shift, about ~45º, of the minimum in $R_{Py}(\theta_a)$ of the (a) and (b) curves simply reflects that the exchange bias field has been pinned at different angles when FC being at $\theta_{FC} = 0^o$ and for $\theta_{FC} = 45^o$.

It would be tempting to extract $H_{eb}$ from the measured $R_{Py}(\theta_a)$ curves using a simple geometrical model of field addition that emerges from the previous discussion (that we have already used in Ref. 6; see also Ref 7). However, use of this simple model is limited to small variation of the magnetization direction (at small applied fields) when the response is reversible. When hysteresis in the AMR response appears, the model is not valuable. The situation is even more complex when one realizes that the in-plane magnetic anisotropy of these hexagonal manganites leads to strong training effects and eventually to a rotatable anisotropy. In this situation the analysis of the AMR is not longer straightforward [9]. In spite of these difficulties, and aiming only to provide an order of magnitude of $H_{eb}$, we have included the fit (solid lines through the experimental points in Fig. 2) of the $R_{Py}(\theta_a)$ curves using the simplest model [6, 7]. It turns out that $H_{eb}$ = 270±40 Oe, 220±20 Oe, 0.2±0.6 Oe and ~ 0 Oe for (a), (b), (c) and (d), respectively. As we discussed in Ref. 5, extraction of the $H_{eb}$ values from AMR data, in the reversible limit, commonly gives values larger that those extracted from magnetization measurements [6].

We turn now to the electric field effects on $H_{eb.}$ We have measured the AMR response when biasing the Py/$YMnO_3$/Pt sandwich by an electric field ($V_e$). In Fig. 3 we show the data recorded, clockwise, at 5 K ($H_a$ = 50 Oe, after 3 kOe FC from room temperature) at some selected biasing-voltages applied ($V_e$ = ±0.05, ±0.4, ±0.8, ±1.2 and ±1.8 V). Comparison of the $R_{Py}$ ($\theta_a$) curves in Figs. 2 and 3 reveals that upon electric-field biasing the $H_{eb}$ gradually disappears. The suppression of magnetic exchange-bias by electric field of the underlying $YMnO_3$ ferroelectric layer suggests a substantial modification of the antiferromagnetic domain structure which is driven by the electric field.

Reversibility of the AMR upon switching the electric field has been investigated. The AMR data recorded at zero biasing-voltage, after measuring under certain $V_e$ bias-field, were found to be different from the virgin AMR curve, measured just after FC from T > 100 K at $\theta_a = 45^o$. To illustrate this effect, we show in Fig. 4, the clockwise $R_{Py}(\theta_a)$ curves measured at $V_e$ = +0.05 V, +1.8 V and +0.05 V sequentially. From the comparison of the initial and final $R_{Py}(\theta_a)$ curves measured at $V_e$ = +0.05 V, it is clear that the exchange bias-field has changed irreversibly upon application of the electric-biasing field *and* rotation of the sample. Indeed, the minimum of the $R_{Py}(\theta_a)$ curves clearly appears shifted. More precisely, within the accuracy of the model used to fit the data -the solid lines through the experimental data-points correspond to the fits performed using the same model as described above- it turns out that the exchange bias field has not changed in magnitude but it has rotated some $60^o$ from its initial position. Detailed discussion of the implications of this hysteretic behavior is out of the scope of this manuscript, but these data clearly illustrates the effects of the electric field on the magnetic structure of the $YMnO_3$.

Attentive inspection of data in Fig. 3 shows two additional important effects. First, there is a gradual reduction of the measured voltage (up to -1.6 %) along the Py layer evidencing that $R_{Py}$ lowers when $|V_e|$ increases. Second, there is a weak dependence of $R_{Py}$ on the electric field polarity and thus $R_{Py}(+V_e) \neq R_{Py}(-V_e)$. We will sequentially discuss these effects. In Fig. 1, we included a diagram indicating the measuring arrangement used in these experiments. We stress that the current injected to the Py track ($I_{Py}$) has been fixed at 100 μA and thus the decrease of the measured $R_{Py}$, when increasing $|V_e|$, indicates that the real current flowing along Py changes with $V_e$ thus suggesting that the current leakage across $YMnO_3$ ($I_{YMO}$) may play a role. We have measured $I_{YMO}$ (at 5 K and $H_a$ = 50 Oe) as a function of $V_e$ (Fig. 5a). Importantly enough, in the explored voltage region, a clear not-ohmic behavior of $I_{YMO}(V_e)$ is observed. This can be well appreciated in Fig. 5b (left axis) where we plot the differential resistance obtained from data of Fig. 5a. Data in this picture shows that the conductance of the $YMnO_3$ film increases ~30% when increasing $V_e$ up to 1.8 V. Therefore, it is clear that the current flowing along the Py layer must be reduced and the corresponding voltage measured on the Py film, in agreement with the experimental observation, must be smaller. Therefore, the reduction of the voltage drop measured on the Py layer is due to the non-ohmic nature of the $I_{YMO}(V_e)$ across the $YMnO_3$. By the same token, when changing the polarity of the bias electric field, there should be a concomitant modulation of the observed resistance. This is also in agreement with the experimental observation.

The previous discussion illustrates the critical role of current leakage on the actual performance of the device. We notice that, among others, there is Joule power dissipated within the $YMnO_3$ and thus some local rising of temperature can not be disregarded. The data reported in Fig. 3, 4 and 5 have been collected at 5 K, as indicated by the thermometers of our PPMS system; no significant rising of temperature was observed during experiments and the resistance of the Py films remained unchanged after sample rotation from -5º to +365º and back to -5º. However, the nanometric thickness of the device does not allow concluding, on solid experimental grounds, that the device temperature remained fixed at the nominal temperature of the experiments (i.e. 5 K). In spite of this, we notice in Fig. 2-c that the exchange bias in Py/$YMnO_3$/Pt is well visible up to 50 K (at least), whereas it is almost washed out at 5 K by electric-field biasing 1.8 V. Additionally, in Fig. 3 it can be also appreciated that the $R_{Py}(\theta_a)$ curves measured using different field polarities gradually split up; i.e the difference $\Delta R_{Py} = R_{Py}(\theta_a, +V_e) - R_{Py}(\theta_a, -V_e)$ depends on $|V_e|$. In Fig. 6 we depict the difference $\Delta R_{Py}$ evaluated at a given $\theta_a$ value (181º). Similar plots are obtained at any other angle. This $\Delta R_{Py}$ dependence on $V_e$, particularly the observed change of sign of $\Delta R_{Py}$ at ~ 0.4 V, would not be expected if heating effects were relevant. Therefore, the reported data strongly suggest a genuine electric-field effect on the exchange bias. This conclusion is also supported by the recent observation of magnetization switching by the electric field [6]. Indeed, it has been shown that the evolution of magnetization under cycling an electric field, particularly the reduction of the magnitude of the magnetization when retreating the electric field, could not be explained by heating effects [6].

In summary, we have shown that an electric field can be used to control the anisotropic magnetoresistance of exchange-biased heterostructures using AF&FE materials and to monitor the modification of the exchange bias. Owing to the fact that electric and magnetic domains are coupled in $YMnO_3$, we consider that the modification of exchange bias by an electric field can be related to the modification of the magnetic domain structure and particularly the domain wall configuration, when biasing the device with an electric field. Due to the low in-plane magnetic anisotropy inherent to the triangular and frustrated antiferromagnetic nature of the hexagonal $YMnO_3$ structure, the magnetic exchange bias rapidly decays when increasing temperature and therefore the electric losses in the ferroelectric $YMnO_3$ can play a substantial role in the measured response. Therefore, beside optimization of losses in these $YMnO_3$-based heterostructures, it is clear that biferroic antiferromagnets with stronger magnetic anisotropy could be more appropriate for optimal functionalization of these devices.

Acknowledgments. Financial support by the MEC of the Spanish Government (projects NAN2004-9094-C03 and MAT2005-5656-C04) and by the European Union (project MaCoMuFi (FP6-03321) and FEDER) are acknowledged.

## Figure captions

**Figure 1.** Sketches of the contact configuration -lateral view- for electric measurement (top panel) and anisotropic magnetoresistance measurements -planar view- (bottom panel). **J** and $\mathbf{H_a}$ are the current density along Py layer and the applied magnetic field, respectively.

**Figure 2.** Angular dependence of the magnetoresistance of the Py film in applied magnetic field $H_a$, after field-cooling (3 kOe). At 5K, measurements were performed at different cooling angles: $\theta_{FC} = 0^o$, $H_a = 50$ Oe (a) and $\theta_{FC} = 45^o$, $H_a = 50$ Oe (b), whereas at 50 K, $H_a = 40$ Oe (c) and 100 K, $H_a = 50$ Oe (d) data correspond to $\theta_{FC} = 45^o$. Data of (b), (c) and (d) curves have been shifted by 0.011 Ω, -0.031 Ω and -0.1063 Ω, respectively, from each other to avoid overlapping. The solid lines through the experimental data-points correspond to the fits (see text).

**Figure 3.** Angular dependence of the magnetoresistance of the Py film, when biasing the Py/$YMnO_3$/Pt sandwich by an electric field corresponding to indicated biasing voltage ($V_e$). Measurements were done at 5 K in magnetic field of 50 Oe, after field-cooling (3 kOe, $\theta_{FC}= 45^o$).

**Figure 4.** Angular dependence of the magnetoresistance of the Py film when biasing the Py/$YMnO_3$/Pt sandwich at $V_e$ = +0.05 V (curve 1), +1.8 V and back to +0.05 V (curve 2). Measurements were done at 5 K in magnetic field of 50 Oe, after field-cooling (3 kOe, $\theta_{FC} = 45^o$). The solid lines through the experimental data-points correspond to the fits (see text).

**Figure 5.** (a) Current across the $YMnO_3$ ($I_{YMO}$) film versus biasing voltage measured at 5 K. Dashed line indicates the linear (ohmic) behavior. (b) Biasing-voltage dependence of differential resistance across the $YMnO_3$ film evaluated from raw data in Fig. 5a (left axis) and resistance $R_{Py} \equiv V_{Py}/I_{Py}$ measured along the Py film at 5 K at $\theta_a = -5^o$ and negative biasing (right axis).

**Figure 6.** Biasing-voltage dependence of the difference between $R_{Py}(\theta_a)$ curves of Fig. 3, measured under electric fields of different polarity: $(\Delta R_{Py} = R_{Py}(\theta_a,+V_e) - R_{Py}(\theta_a,-V_e))$. Data is reported for $\theta_a = 181^o$.

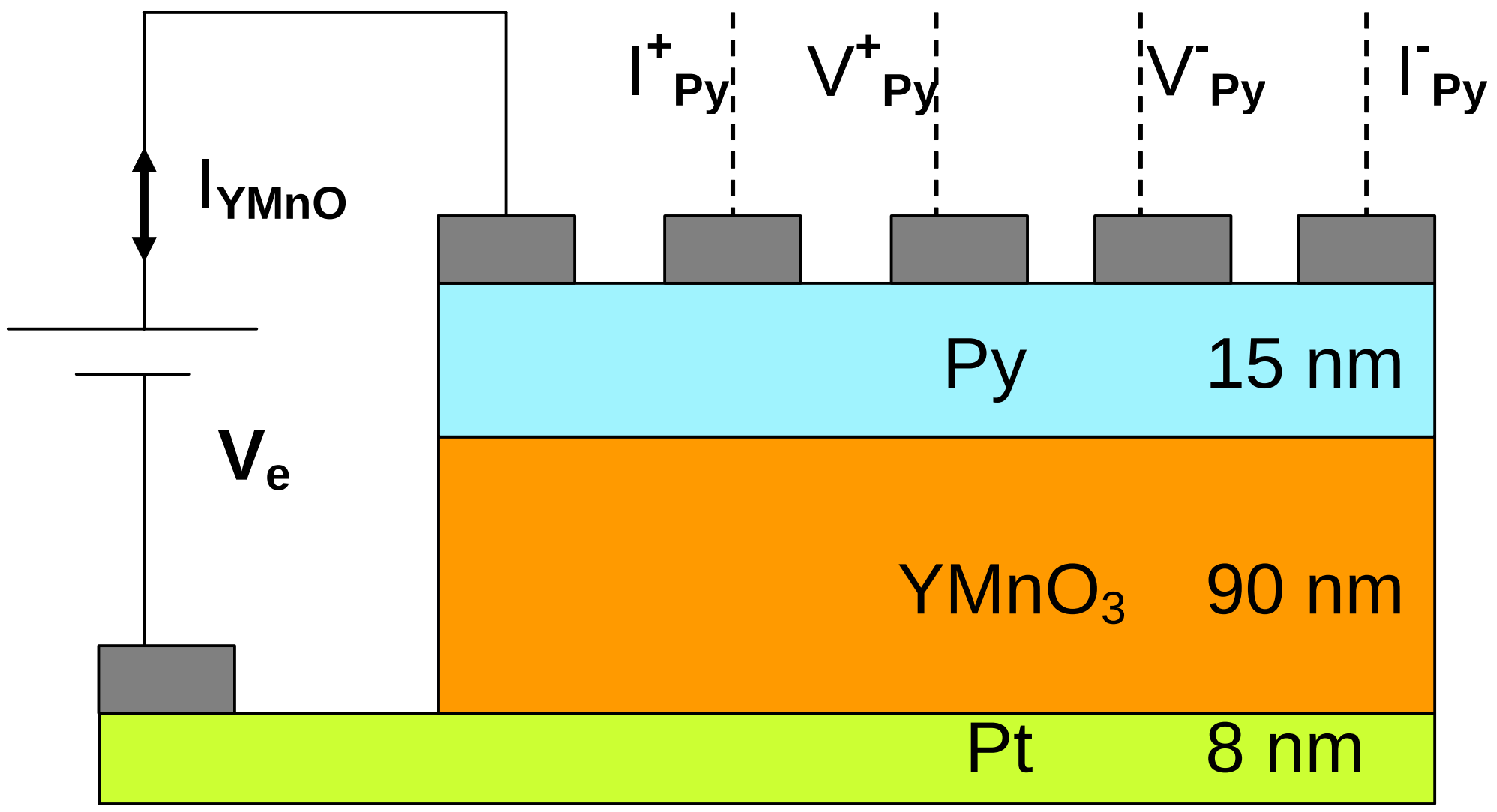

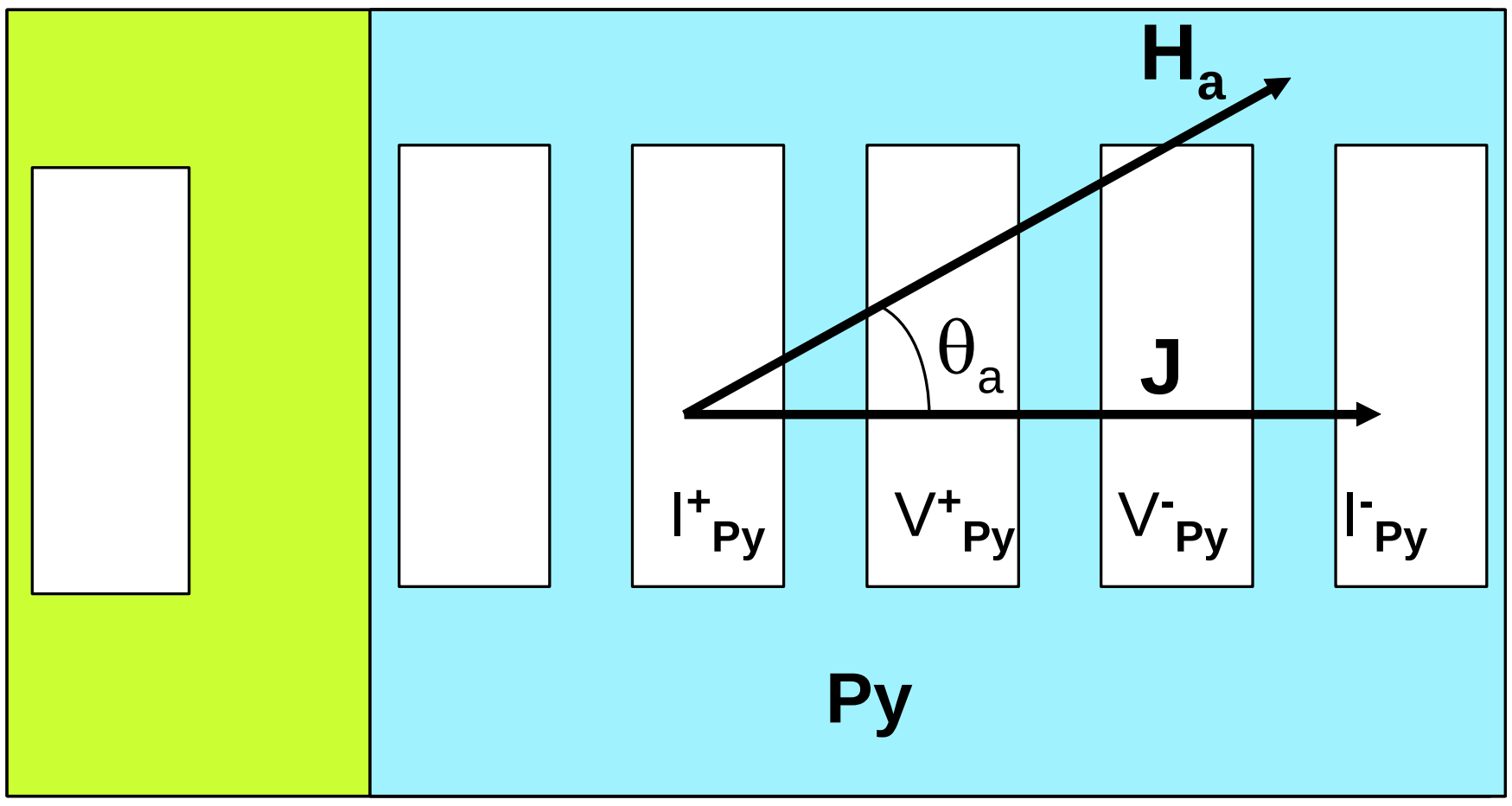

**Fig. 1**

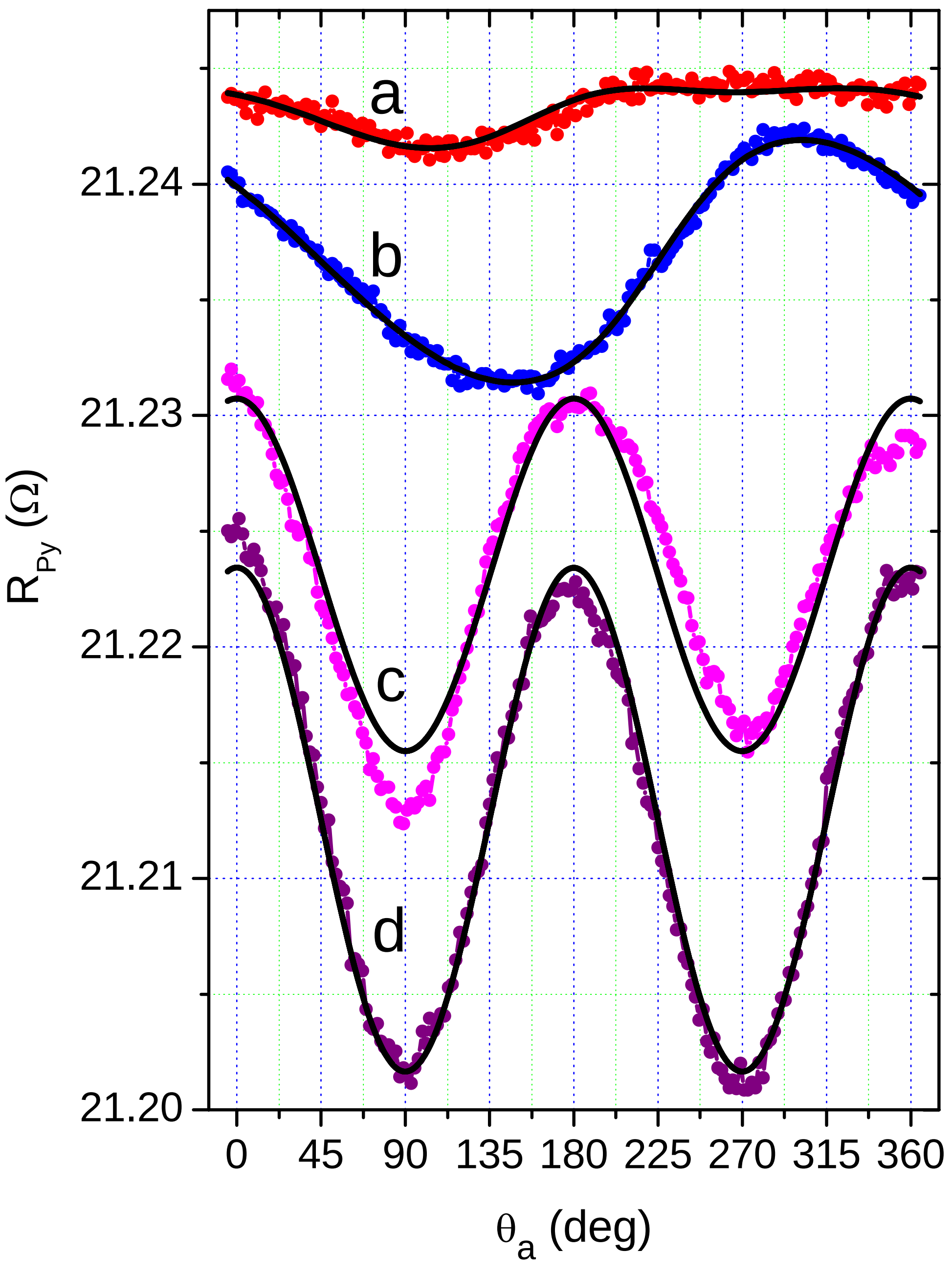

**Fig 2**

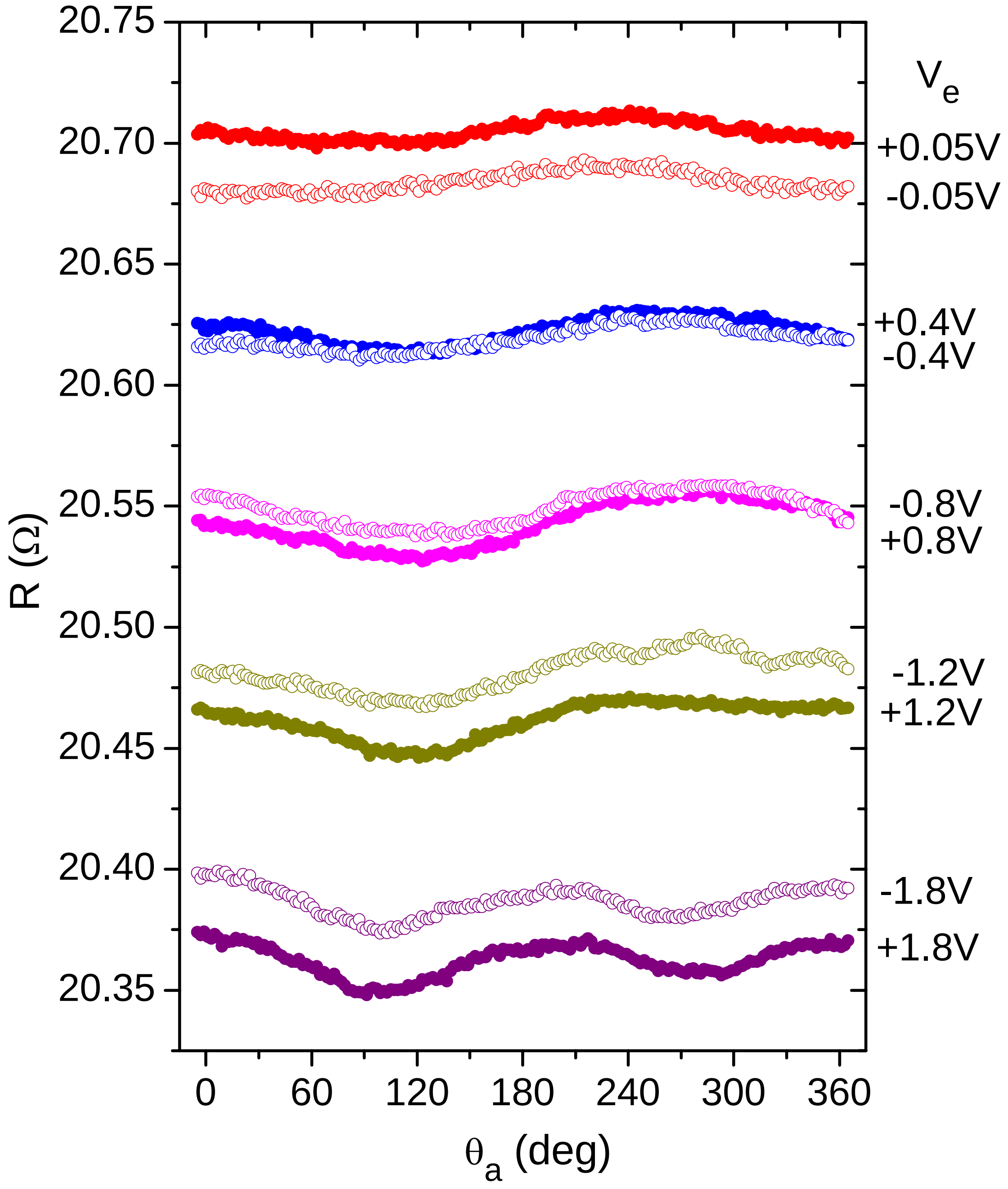

**Fig 3**

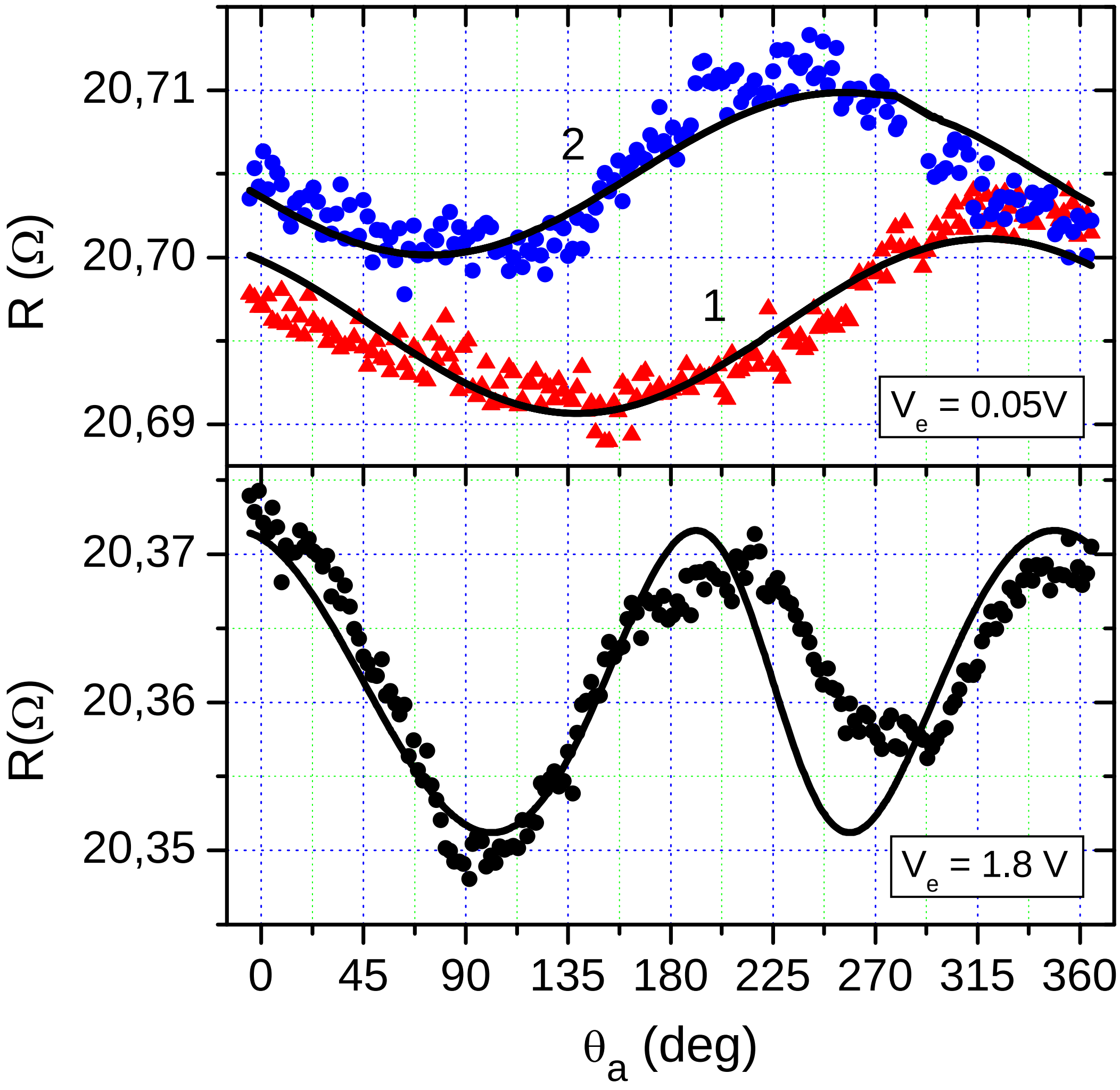

**Fig 4**

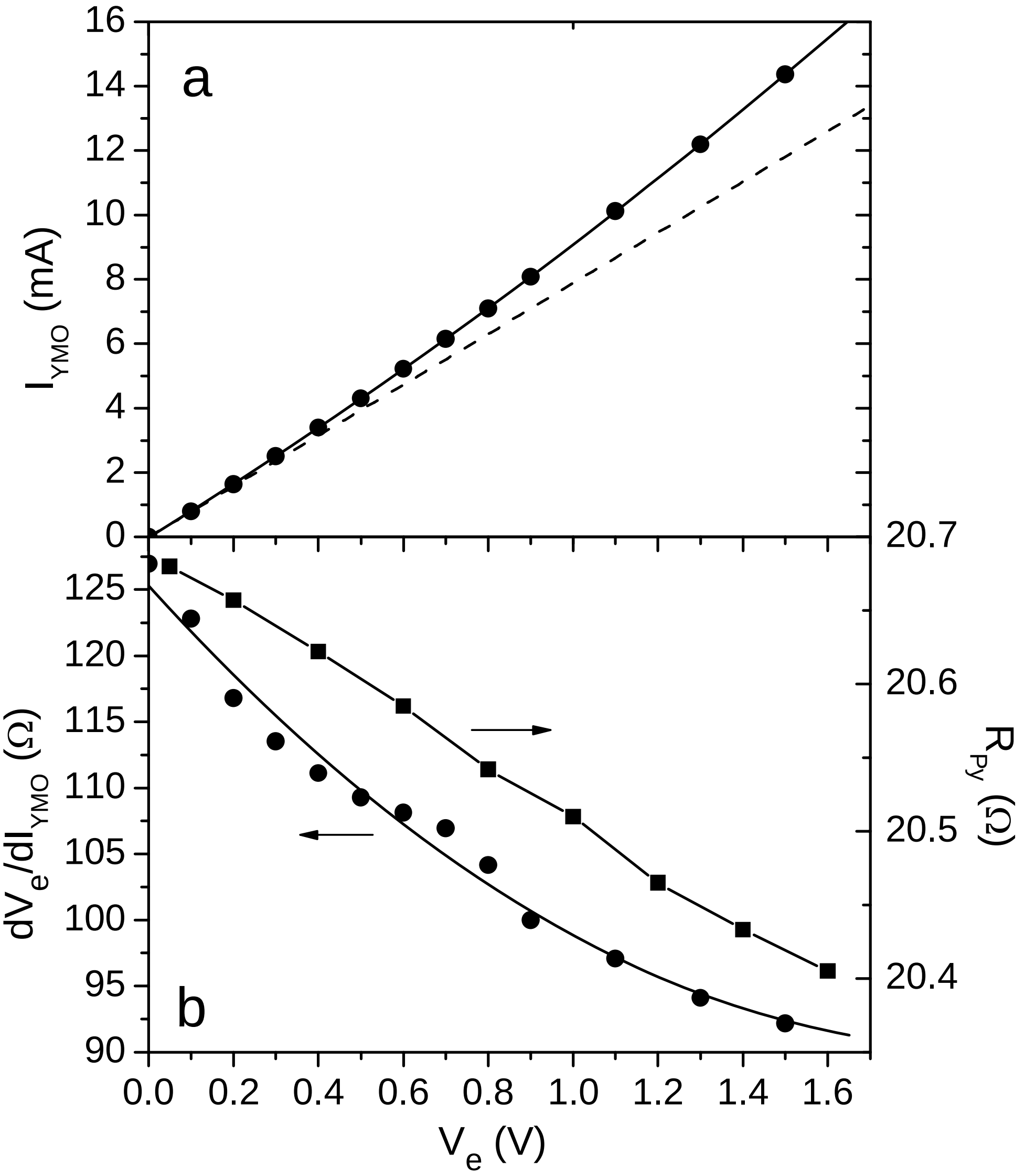

**Fig 5**

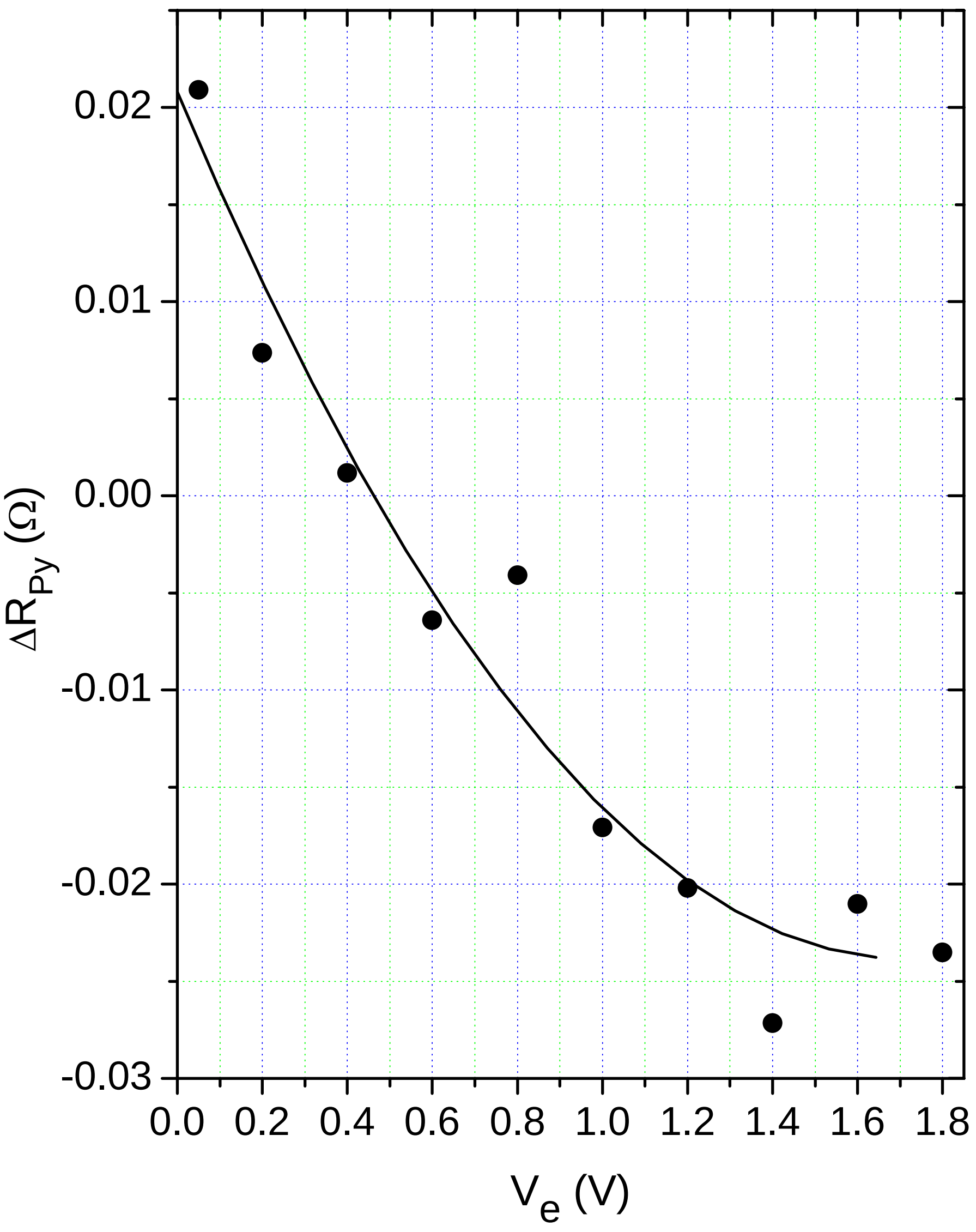

**Fig. 6**